\documentclass[aps,pre,superscriptaddress]{revtex4}

\usepackage{graphics}      
\usepackage{graphicx}      
\usepackage{longtable}     
\usepackage{url}           
\usepackage{bm}            
\usepackage{amsmath}
\usepackage{amssymb}
\usepackage{wasysym}
\usepackage{color,soul}
\usepackage{ulem} 
\usepackage{subfigure}
\usepackage{bbold}

\usepackage{amsmath}
\usepackage{amssymb}
\usepackage{subfigure}
\usepackage{hyperref}
\usepackage{url}
\usepackage{xcolor}
\usepackage{color}
\definecolor{amaranth}{rgb}{0.9, 0.17, 0.31}
\definecolor{purple(munsell)}{rgb}{0.62, 0.0, 0.77}
\definecolor{americanrose}{rgb}{1.0, 0.01, 0.24}
\definecolor{palatinateblue}{rgb}{0.15, 0.23, 0.89}
\definecolor{royalblue(web)}{rgb}{0.25, 0.41, 0.88}
\definecolor{hanpurple}{rgb}{0.32, 0.09, 0.98}
\definecolor{beaublue}{rgb}{0.74, 0.83, 0.9}
\definecolor{carminered}{rgb}{1.0, 0.0, 0.22}
\definecolor{brightpink}{rgb}{1.0, 0.0, 0.5}
\definecolor{vividviolet}{rgb}{0.62, 0.0, 1.0}
\hypersetup{ linktoc=all,
	colorlinks, linkcolor={palatinateblue},
	citecolor={brightpink}, urlcolor={royalblue(web)}}

\begin{document}

\title       {Entropic analysis of the localization-delocalization transition in a one-dimensional correlated lattice}
\author{O. Farzadian}
\affiliation{Department of Physics, School of Science and Technology, Nazarbayev University, Astana 010000, Kazakhstan}

\author{T. Oikonomou}
\email{thomas.oikonomou@nu.edu.kz}
\affiliation{Department of Physics, School of Science and Technology, Nazarbayev University, Astana 010000, Kazakhstan}

\author{M. R. R. Good}
\affiliation{Energetic Cosmos Laboratory, Nazarbayev University, Astana 010000, Kazakhstan}

\author {M. D. Niry}
\affiliation {Department of Physics, Institute for Advanced Studies in Basic Sciences (IASBS), Zanjan 45137-66731, Iran}
\affiliation {Center for Research in Climate Change and Global Warming (CRCC), Institute for Advanced Studies in Basic Sciences (IASBS), Zanjan 45137-66731, Iran}

\date{\today}

\begin{abstract}
In this work, propagation of acoustic waves in a one-dimensional binary chain with different types of correlations in elasticity distribution is studied. We applied entropic analysis to investigate and quantify the localization-delocalization transition in long-range correlated chains in terms of the scaling exponent $\alpha$ and discuss its relation to the order-disorder levels in the structure of the chain. The results demonstrated that the entropic consideration detects correctly the critical value of $\alpha$ separating localization from delocalization bands. 
\end{abstract}

\maketitle


\section{\label{sec:Int}Introduction}

Currently, Anderson localization transition is well-understood in one-dimensional (1D) random media \cite{Anderson1979,Figotin1996,Figotin1997,Barros2011,Costa2011}. According to the scaling theory of localization \cite{Anderson1979}, in 1D systems, any weak white noise disorder will localize all the (electron) states \cite{Anderson1958}. This means that the envelope of the wave function $\Psi(r)$ decays exponentially in the spatial domain, $\Psi(r)\propto\exp(-r/\xi)$, where $\xi$ is the localization length. On the other hand, a variety of studies show that the presence of correlated disorder, even in the 1D Anderson tight binding model, can generate delocalized states \cite{Dunlap1990,Moura1998,Flores1089,Izrailev1999,Carpena2002, Diez1994, Bellani1999}. The relevance of the underlying long-range correlations to the electronic transport in a 1D chain was first explained by Carpena {\it et al.} \cite{Carpena2002}. They numerically showed that for a binary chain exhibiting long-range correlations (LRC) with a scaling exponent $\alpha$ greater than a threshold value ($\alpha_c\sim1.45$), there is a broad energy band of extended (delocalized) states. Alternatively, the extended states can be produced by the presence of short-range correlations  (SRC) in binary random sequences  \cite{Barros2011,Farzadian2016,Farzadian2018}, e.g., the random-dimer model \cite{Costa2011,Esmailpour2008}. Consequently, short- and long-range correlations are used to examine the localization-delocalization transition in disordered chains \cite{Farzadian2016, Barros2011, Costa2011, Moura1998, Esmailpour2008}. Also, heat conduction in 1D harmonic chains with correlated disorder has long been a subject of investigation \cite{Dhar2008,Lepri2003}. Analytical and numerical results demonstrate that specific long-range correlations of isotopic disorder can suppress or enhance the heat flow through disordered chains \cite{Herrera2010}.

In this work, we use a simple 1D bead-spring model, which consists of two distinct spring constants (denoted also as elasticity of the chain) $k_\mathrm{A}$ and $k_\mathrm{B}$. Their distribution on the chain is dictated by processes exhibiting SRC ($ n $-mers) or LRC for numerous values of the scaling exponent $\alpha$. In the first part of our study we apply the Transfer Matrix method on the wave equation which describes the pulse propagation in the lattice, to calculate the localization length $\xi$ as a function of the wave frequency $\omega$. For SRC in the elasticity distribution we numerically demonstrate the existence of discrete extended modes  which appear as peaks in the function $\xi(\omega)$. For frequencies different than these resonance frequencies there is no wave propagation. A qualitatively different picture is recorded for elasticity distributions with LRC features. Namely, depending on the value of $\alpha$, one observes either a localized or an extended band (continuous resonance frequency band) for the entire frequency spectrum. 

In the second and main part of our work, we apply entropic analysis as a measure of complexity to analyze and quantify the localization-delocalization transition in our lattice.
For this, we invoke two types of entropic measures, the spectral entropy \cite{Inouye1991} and the block entropy \cite{Nicolis}, both based on the Shannon probabilistic expression $\sum_i p_i \ln(1/p_i)$. Various other entropic functionals have been used in the literature to capture transition phenomena \cite{Marzolino2013,Stephan2011,Misra2015,Helmes2015}. However, on the basis of the fulfillment  of the Shore-Johnson axioms for a proper inference procedure \cite{Oikonomou2019, Thomas2018}, we chose the Shannon expression.
Particularly, we generated several time series exhibiting LRC, $\alpha\in(1,2)$, in the elasticity distribution and calculated their power spectra. After normalization we studied the respective spectral entropy as a function of the scaling exponent $\alpha$ for several sequence lengths $N$. Mapping the power spectra of large size $N$ (to avoid finite size effects) into  binary sequences we considered the aforementioned transition phenomenon at the level of the sequences' internal structure, by applying the block entropy analysis and exploring its behavior again with respect to $\alpha$ for a fixed $N$. Both types of complexity measures were able to detect  the correct threshold value of $\alpha$, namely $\alpha=\alpha_c$ as predicted by the theory. Even more interesting was the observation that both measures exhibited the same qualitative picture of the entropy dependence on the scaling exponent.

The paper is organized as follows: In Sec.~\ref{sec:Model}, we introduce the general model and provide a solution which allows us to calculate the localization length $\xi$ as a function of the wave frequency $\omega$. In Sec.~\ref{sec:Numerical} the numerical results of our model are presented. The Shannon (spectral and block) entropy as a measure of complexity of the chain is invoked to describe and quantify the localization-delocalization transition in Sec.~\ref{sec:Entropy}. Finally, a summary of the results is presented in Sec.~\ref{sec:Dis}.

\section{\label{sec:Model}Model and Methods}

Consider a scalar wave equation in a 1D lattice with random distribution of the elastic constants, 
\begin{eqnarray}\label{eq:Scalarwave}
\frac{\partial^2}{\partial t^2}\phi(x,t) = \frac{\partial}{\partial x} \left[k (x) \frac{\partial}{\partial x} \phi (x,t)\right],
\end{eqnarray}
where $\phi(x,t)$ is the wave function, $x$ and $ t $ denote the spacial and temporal dependence, respectively, and $k (x)$ denotes spring constants (the mass is set to unity throughout the paper). The wave function is of the form $\phi (x,t)= \psi (x)\exp(-\mathrm{i} \omega t)$, where $\omega$ is the wave frequency and $\psi(x)$ is the wave amplitude. We use the Finite Difference Method (FDM) to write the wave equation in its discretized form. By setting the nearest-neighbor spacing, $\Delta x=1$, the wave equation for the site $i$ on the chain, is obtained as
\begin{equation}\label{eq:discrete}
-\omega^2 \psi_i = k_i(\psi_{i+1}-\psi_{i}) - k_{i-1}(\psi_{i}-\psi_{i-1}).
\end{equation}
To investigate the propagation of the wave, Eq.~(\ref{eq:discrete}) can be expressed in terms of the conventional Transfer Matrix method by the following recursive
matrix form:
\begin{eqnarray}
\left(\begin{array}{c}
\psi_{i+1}\\
\psi_{i}
\end{array} \right) =M_{i,i-1}\left(\begin{array}{c}
\psi_{i}\\
\psi_{i-1}
\end{array}\right),
\end{eqnarray}
where
\begin{eqnarray}\label{eq:MT}
\label{matrix}
M_{i,i-1} = \left(\begin{array}{cc}
\frac{-\omega^2 + k_i + k_{i-1}}{k_i} & -\frac{k_{i-1}}{k_i} \\
\\
1 & 0
\end{array}\right)\,.
\end{eqnarray}
The wave amplitude at the $(N+1)$th site of the chain is related to its value at the first site by calculating a product of matrices $Q_{N,1}= \prod_{i=1}^{N} M_{i,i-1}$ as
\begin{eqnarray}
\left(\begin{array}{c}
\psi_{N+1}  \\
\psi_{N}
\end{array} \right) = Q_{N,1} \left(\begin{array}{c}
\psi_1 \\
\psi_0
\end{array}\right),
\end{eqnarray}
where the initial value is $\psi_1=\psi_0=1/\sqrt{2}$. The evolution of the wave function is determined by the multiplication of  consecutive transfer matrices. 
After multiplication of a limited number $\tau$ of matrices, the amplitude of the wave function describing localized states becomes very large. 
We normalize the achieved wave function and save $d=d(\omega)$ as the normalization factor to avoid numerical errors. A new wave function is obtained by multiplication of $d$ with  the magnitude of the previous wave function. After the $m$th normalization we calculate the Lyapunov exponent  defined by
\begin{equation}\label{eq:Lyapunov}
\gamma(\omega)=\frac{1}{m\tau}\sum_{r=1}^{m}\ln\big({d_r(\omega)}\big)\,,
\end{equation}
where $d_r$ denotes the normalization factor $d$ of the $r$th normalization process. The inverse of the Lyapunov exponent yields the localization length, $\xi(\omega) = 1/\gamma(\omega)$ \cite{Anderson1979}.

The calculation of the localization length by means of Eq.~(\ref{eq:Lyapunov}), as a self-averaging quantity, requires a long chain.
As such, our study is conducted in terms of the wave frequency $\omega$ for a fixed chain of length $N=2^{18}$.
We generate a binary chain consisting of two elastic constants $k_\mathrm{A}$ and $k_\mathrm{B}$. Their distribution in the medium is dictated by the nature of the correlations we are implementing in the chain. 

To generate the long-range correlated sequences regarding the $k_i$'s, the modified Fourier filter method \cite{Stanley1996} is applied. Then, we map every positive and negative value of the correlated series into springs $\mathrm{A}$ and $\mathrm{B}$, respectively \cite{Farzadian2016,Carpena2002}. These sequences are used as the $k_i$'s in the binary chain. Such mapping can change the correlation properties of the original sequence. In order to quantify the correlation in the final binary sequence, we calculate the scaling exponent $\alpha$ by using Detrended Fluctuation Analysis (DFA) as described in \cite{Hu2001,Chen2005}. 
For generating sequences characterized by SRC,  $k_\mathrm{B}$ is chosen as the reference spring constant. Next, we randomly locate in the chain $n$ repetitions of the $\mathrm{B}$ spring, the so-called $n$-mers.
A characteristic example is a dimer ($2$-mer) chain with respect to $k_\mathrm{B}\to B$ as, $\mathrm{AAA(BB)AA(BB)(BB)A\ldots}$, in which the $k_\mathrm{B}$ values appeared only in pairs of neighboring sites of the chain.  
Without loss of generality of our results, we randomly substituted numerical values into the elasticity constants as $k_\mathrm{A}=10$ and $k_\mathrm{B}=8$. 

\section{\label{sec:Numerical}Numerical results}

To obtain the localization length $\xi(\omega)$ for two different correlation types in the binary chain, we can follow the proposition introduced in \cite{Farzadian2016,Zhang2005, Esmailpour2008}. Regarding Eq.~(\ref{eq:MT}), there are four transfer matrices related to the $\mathrm{A}$ and $\mathrm{B}$ springs in the lattice, i.e.,
\begin{subequations}
	\begin{eqnarray}
	\label{eq:MAA}
	M_\mathrm{AA}&=&\left(\begin{array}{cc}
	\frac{-\omega^2 + 2k_\mathrm{A}}{k_\mathrm{A}} & -1 \\
	\\
	1 & 0
	\end{array}\right)\,,\\
	\label{eq:MCG}
	M_\mathrm{AB}&=&\left(\begin{array}{cc}
	\frac{-\omega^2 + k_\mathrm{A} + k_\mathrm{B}}{k_\mathrm{B}} & -\frac{k_\mathrm{A}}{k_\mathrm{B}} \\
	\\
	1 & 0
	\end{array}\right)\,,
	\end{eqnarray}
\end{subequations}
and $M_\mathrm{BB}$, $M_\mathrm{BA}$ which are obtained by carrying out the transformations $\mathrm{A} \rightarrow \mathrm{B}$ and $\mathrm{B} \rightarrow \mathrm{A}$ in Eqs. (\ref{eq:MAA}) and (\ref{eq:MCG}), respectively.

We can categorize the correlations in the appearance of $\mathrm{A}$ and $\mathrm{B}$ springs in the generated binary sequences as follows:

\begin{itemize}
	\item SRC in $n$-mer chain which consists of clusters of $n$-times repeated $k_\mathrm{B}$ spring constants.
	\item LRC with $\alpha < \alpha_c=1.45$,
	\item LRC with $\alpha > \alpha_c=1.45$,
\end{itemize}
We recall  that the values of the scaling exponent $\alpha \lessgtr 1$ correspond to the fractional Brownian noise (fBn, $>$) and the fractional Gaussian noise (fGn, $<$), while $\alpha = 0.5$ corresponds to an uncorrelated random sequence (delta correlated Gaussian white noise) \cite{Carpena2002}. As was mentioned previously, localization/delocalization of waves in a disordered medium can be affected by the spatial correlation of the disorder. According to Carpena \textit{et al.}, for $\alpha > \alpha_c=1.45$, we expect to observe an extended band \cite{Carpena2002}.

\begin{figure}[hbtp]
	\centering
	\subfigure{\includegraphics[width=0.47\textwidth]{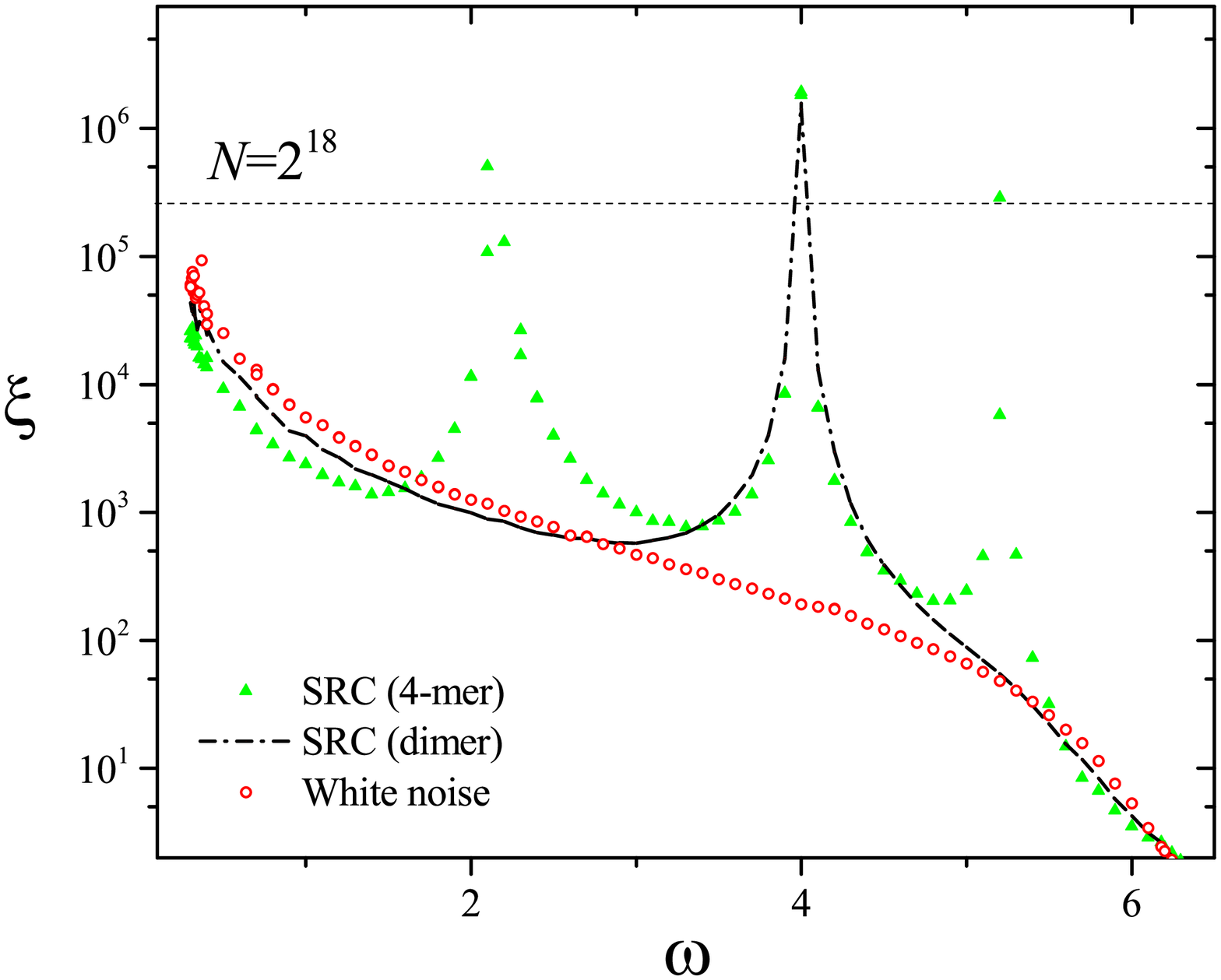}}
	\put(-21,135){$(\mathrm{a})$}
	\quad
	\subfigure{\includegraphics[width=0.48\textwidth]{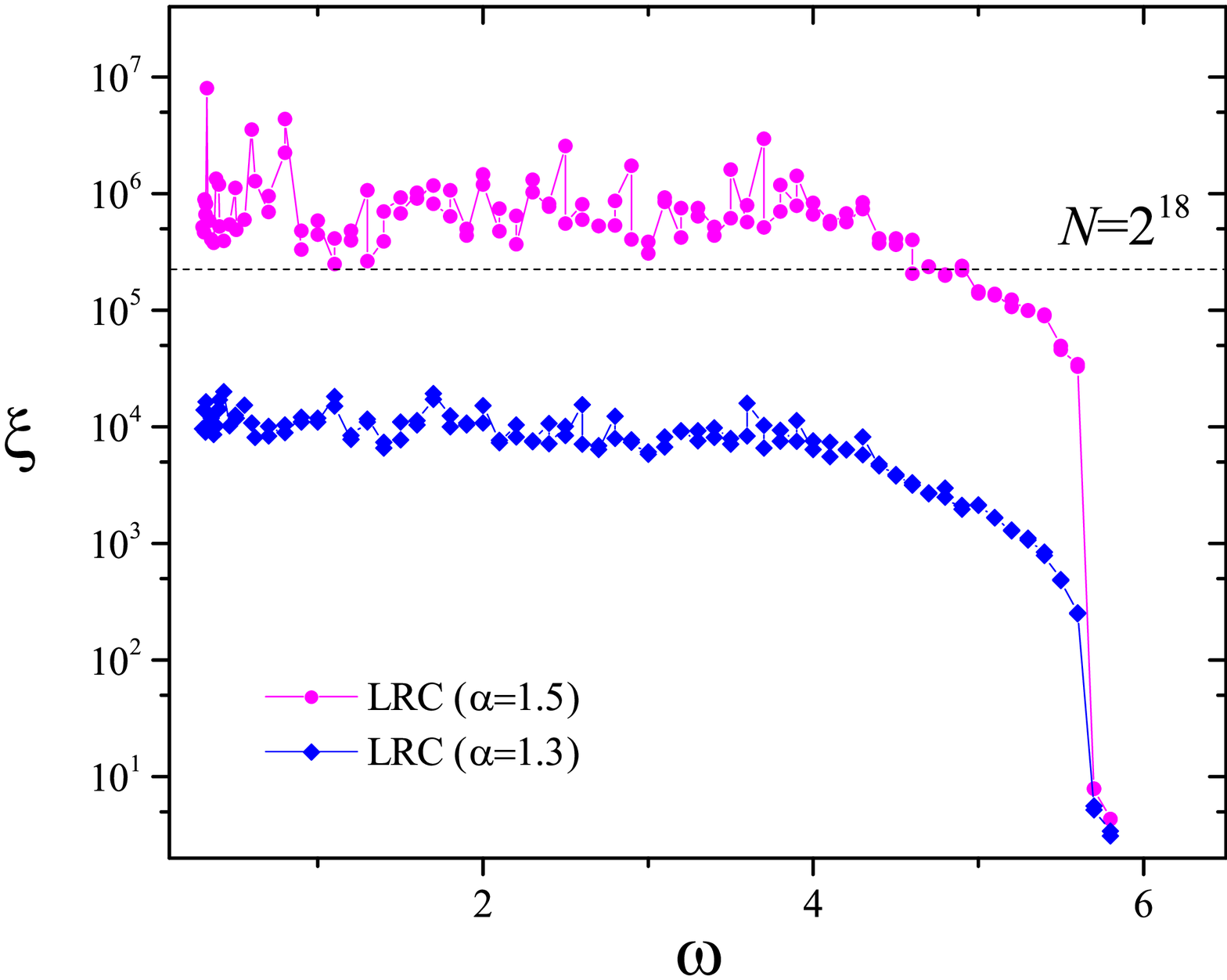}}
	\put(-21,135){$(\mathrm{b})$}
	\caption{ The localization length $\xi$ of a typical short- and long-range correlated chain in terms of the wave propagation frequency $\omega$. $(\mathrm{a})$ The localization length for random $n$-mer chains presented. For a $n$-mer sequence, $(n-1)$ discrete extended modes exist. In a perfectly uncorrelated chain (white noise) there is no extended mode, all frequencies are localized. $(\mathrm{b})$ Two typical long-range correlated chains with different scaling exponents $\alpha=1.3<\alpha_c<\alpha=1.5$. Extended band in exists only for $\alpha>\alpha_c$.}
	\label{fig:lg}
\end{figure}

In Fig.~\ref{fig:lg}$(\mathrm{a})$ the localization length $\xi(\omega)$ is recorded for SRC random $n$-mer chains and a disordered chain generated by a perfectly delta correlated white noise. As expected, for $n$-mer sequence we observe $(n-1)$ discrete extended modes \cite{Farzadian2016}. For example, the $2$-mer chain (dimer, black dashed-dotted line) exhibits one and the $4$-mer chain (green triangles) exhibits three extended modes. Finally, in the case of white noise (random sequence, open red circles) all frequencies are localized.
Similarly, in Fig.~\ref{fig:lg}$(\mathrm{b})$ we plot the localization length as a function of the wave frequency for chains exhibiting LRC ($1<\alpha<2$). For this we considered two LRC chains, associated with a scaling exponent less ($\alpha=1.3$, blue diamonds) and higher ($\alpha=1.5$, purple circles) than the critical value $\alpha_c$. In the former case, we observe that all modes are localized, while in the latter case an extended band of delocalization frequencies appears.

\section{\label{sec:Entropy}Entropy of the Sequence}
Entropy analysis provides a measure of the irregularity or randomness degree within a series of data \cite{Gao2008,Paninski2003,Fell1996401}. In this section we introduce the entropy as a measure of disorder to control the localization-delocalization transition. For this, we invoke the well-known Shannon Entropy (SE) \cite{Shannon1948} defined by
\begin{eqnarray}
\label{Shanonentropy}
SE= -\sum_{i} P_i\log P_i,
\end{eqnarray}
where $P_i=P(x_i)$ is the occurrence probability of the random variable $x_i$.

\subsection{\label{subsec:Shanon} Spectral Entropy}

The entropy measure in Eq.~(\ref{Shanonentropy}) can be calculated by using the spectral method \cite{Fell1996401}. This means finding $F(\omega)$ which is the Fast Fourier Transform of our data series. We then calculate  the associated power spectrum defined as $S_p(\omega)\sim |F(\omega)|^2$ \cite{Fell1996401}. The power spectrum of a signal,  $S_p(\omega)d\omega$, denotes the occurrence probability of the $\omega$ mode in that signal. Using this spectral measure for a discrete time series we can define the probability
\begin{eqnarray}\label {extensive}
P_i(N)=\frac{\sum_{j=1}^{K} S_{p,j}(\omega_i ) }{\sum_{i=1}^{N}\sum_{j=1}^{K} S_{p,j}(\omega_i )}\,.
\end{eqnarray}
The index $j$ runs over $K$ possible chain realizations. Substituting  Eq.~(\ref{extensive}) into Eq.~(\ref{Shanonentropy}) we obtain the Shannon spectral entropy $SE_\alpha(N)$ as a function of the length of the chain $N$.  One can introduce an intensive entropy by dividing the entropy of the chain by the logarithm of the chain size \cite{Smith2007, Fell1996401, Inouye1991}.

\begin{figure}[ht]
	\centering
	\subfigure{\includegraphics[width=0.47\textwidth]{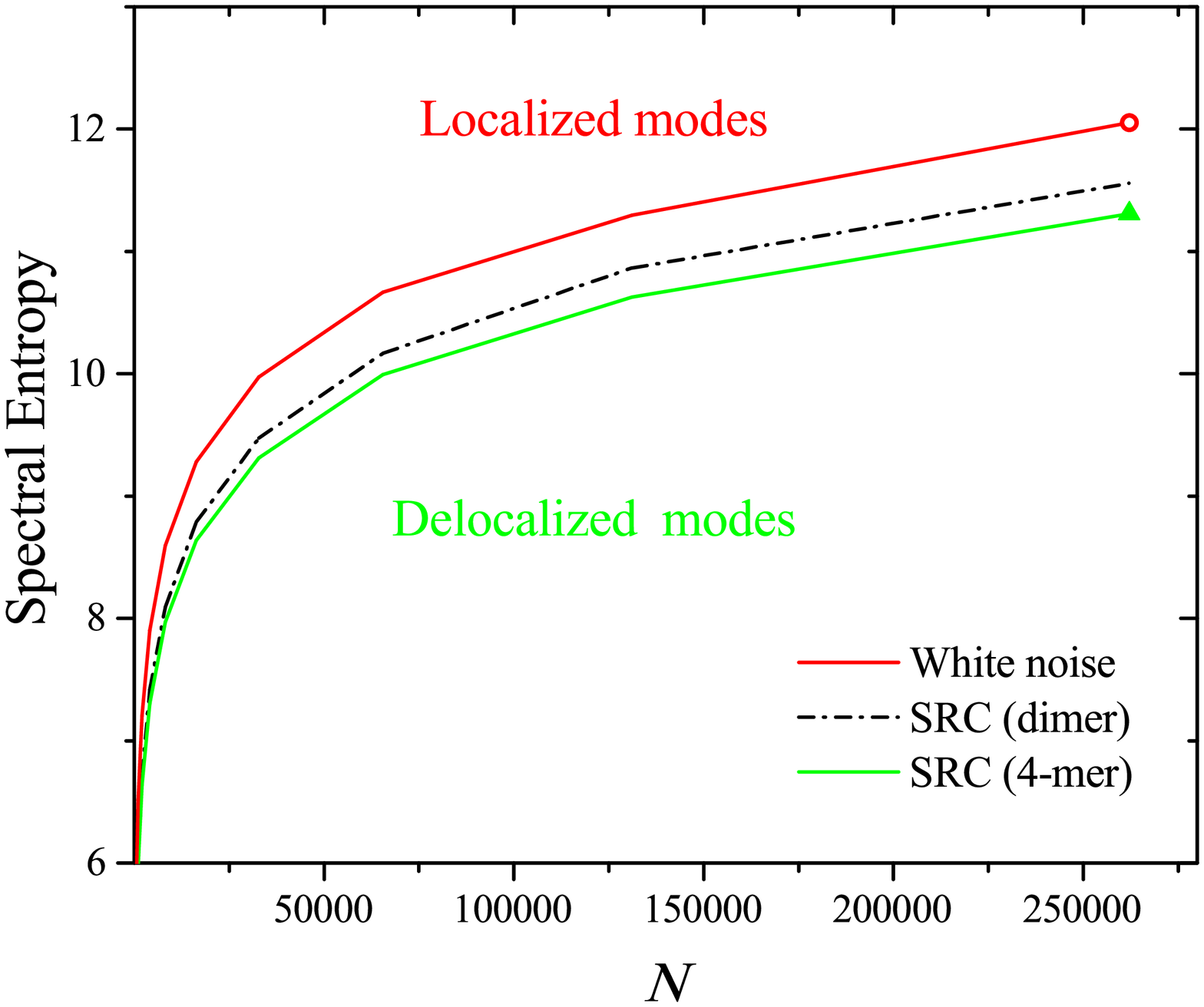}}
	\put(-155,135){$(\mathrm{a})$}
	\quad
	\subfigure{\includegraphics[width=0.47\textwidth]{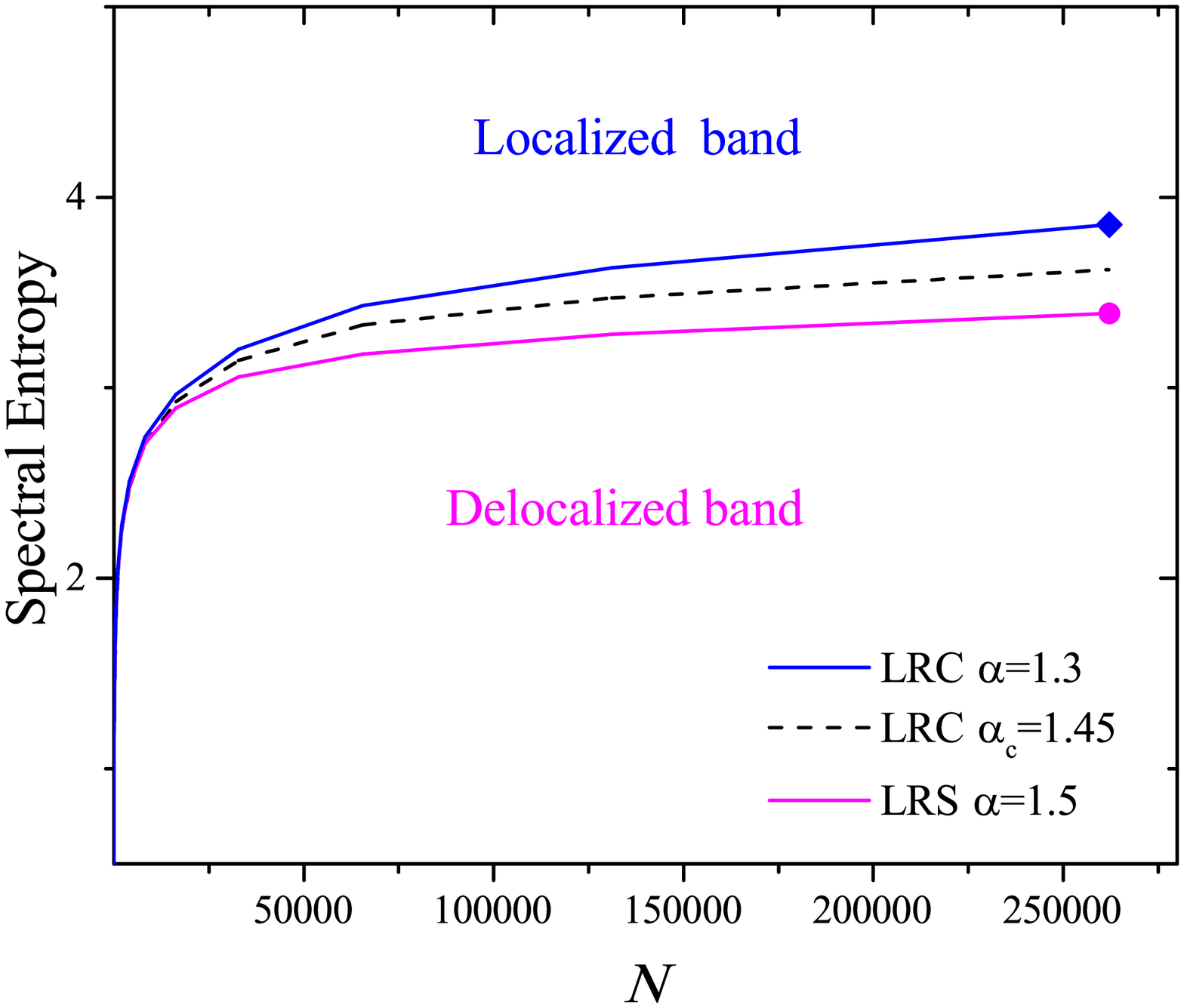}}
	\put(-155,135){$(\mathrm{b})$}
	\caption{\label{fig:entropy} Plot of the spectral entropy as a function of the chain length $N$ for the (a) SRC and (b) LRC sequences presented in Fig.~\ref{fig:lg}. The dashed-dotted line shows the threshold entropy for SRC sequences, above which all states are localized. The dashed line shows the threshold entropy (with $\alpha=\alpha_c = 1.45$) for LRC sequences, below which  extended band exists.}
\end{figure}

In Fig.~\ref{fig:entropy} we consider the spectral entropy of the five  sequences recorded in Fig.~\ref{fig:lg} by varying their length $N$. We observe numerically that the spectral entropy of the dimer chain (dashed-dotted line) can be considered as the threshold entropy for SRC chains distinguishing the regimes between localized (white noise sequence, open red circles) and delocalized modes ($n>2$-mers, green triangle).
Regarding the LRC sequences, we plot the spectral entropy  $SE_{1.3}(N)$ (blue diamond) and $SE_{1.5}(N)$ (pink filled circle), together with the entropy of the critical scaling exponent $\alpha_c=1.45$ (dashed line). Here $SE_{1.5}<SE_{\alpha_c}<SE_{1.3}$ for all $N$. This order of the entropy values gives a first indication that the spectral entropy measure may be able to capture the localization-delocalization transition with $SE_{\alpha_c}$ being the threshold entropy value.
\begin{figure}[htp]
	\centering
	\subfigure{\includegraphics[width=0.6\textwidth]{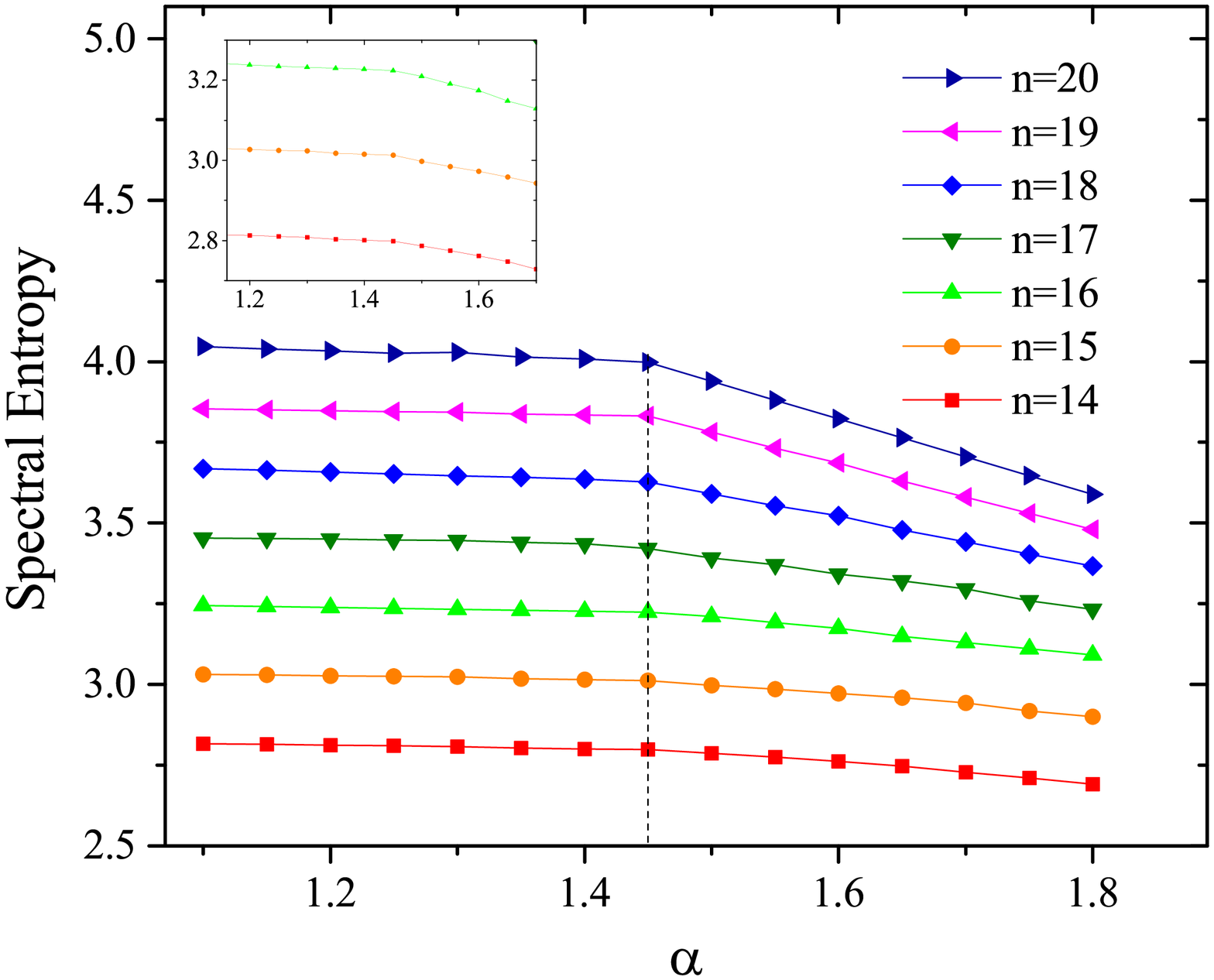}}
	\put(-19,228){$(\mathrm{a})$}
	\quad
	\subfigure{\includegraphics[width=0.6\textwidth]{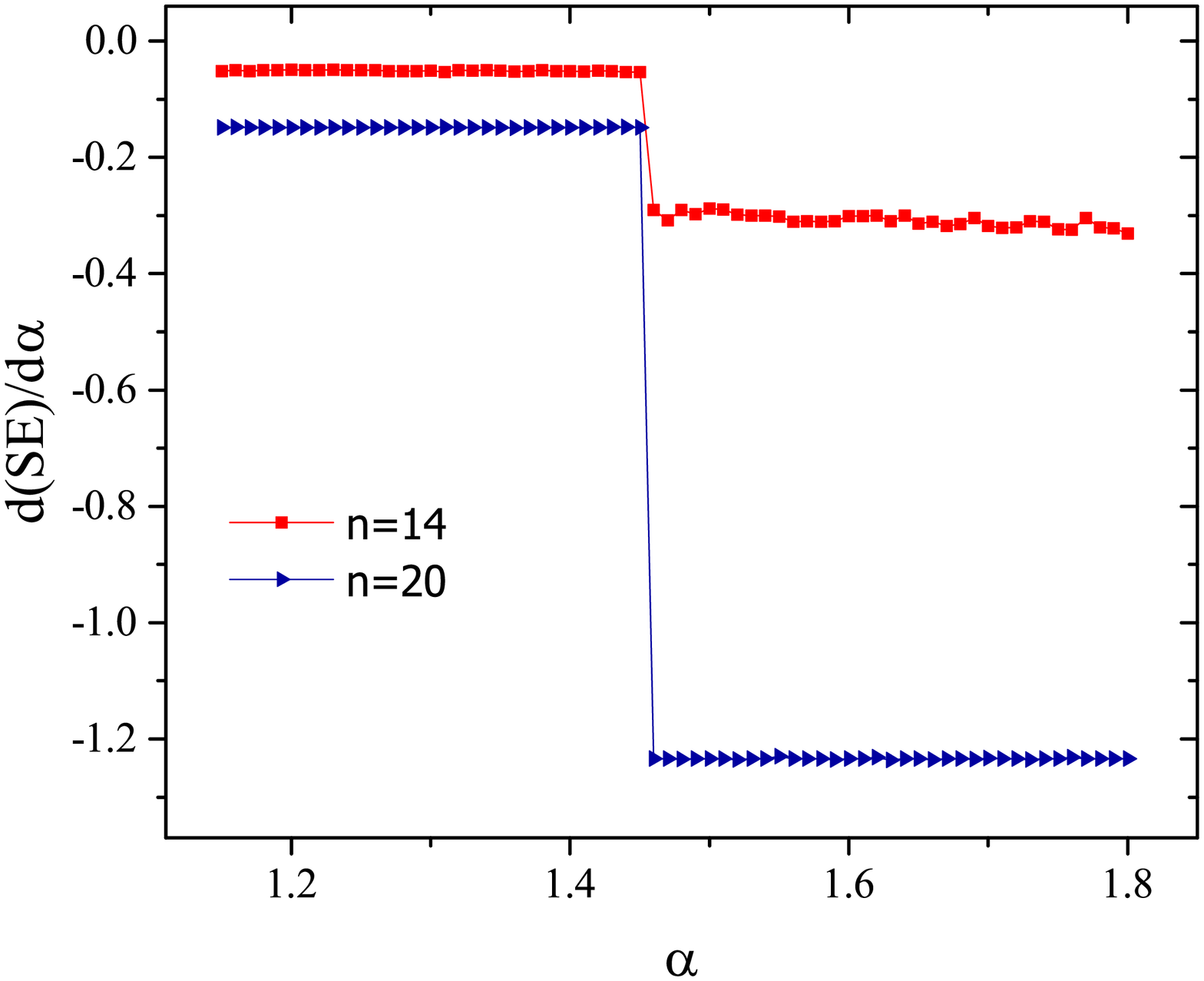}}
	\put(-20,230){$(\mathrm{b})$}
	\caption{ \label{fig:alpha} (a) Plot of the spectral entropy as a function of the scaling exponent $\alpha$ for different LRC chain lengths $N=2^n$. The vertical dashed line corresponds to $\alpha_c$. (b) Plot of the rate of change of the spectral entropy with respect to $\alpha$ in steps of $\Delta \alpha=0.01$.}
\end{figure}

To explore this further, we calculated $SE_\alpha$ as a function of $\alpha$ this time for the LRC regime $\alpha\in[1.1,1.8]$ in steps of $\Delta \alpha=0.05$, by fixing the chain length to $N=2^n$ with $n=14,\ldots,20$ and plotted it in Fig.~\ref{fig:alpha}(a). We observe a  monotone decreasing behavior of $SE_\alpha$ as the scaling exponent $\alpha$ increases for all values of $N$. Precisely, the entropy decays with constant slope up to a certain value of $\alpha$. After this value, entropy decay becomes stepper yet preserves the linear behavior feature. Interestingly enough, this particular value of $\alpha$ below and above which the entropy slope changes is the critical value $\alpha_c$ observed by Carpena \textit{et al}  \cite{Carpena2002} distinguishing the localization-delocalization regimes. 
To ensure that Fig.~\ref{fig:alpha}(a) describes  a transition we plot in Fig.~\ref{fig:alpha}(b) the rate of change of the spectral entropy for the smallest (red square) and the largest (blue triangle) chain size of Fig.~\ref{fig:alpha}(a).  
The discontinuity at $\alpha_c$ becomes apparent. Moreover, we observe that the size of the chain $N$ defines the magnitude of the difference between the different slopes, i.e. for greater $N$ we have greater difference.

\subsection{\label{subsec:Block} Block Entropy}

The Block Entropy $BE(\ell)$ based on the Shannon expression in Eq.~(\ref{Shanonentropy}) is a traditional measure for analyzing the internal complexity of a binary chain. In this case, the probabilities $P(x_i= \ell)$ describe the likelihood of occurrence for a one-dimensional block (or ``word") of size $\ell$ for a fixed length $N=2^{18}$ of the chain with $\sum_{\ell=1}^{\ell_{\max}}P_N(\ell)=1$. The reader may find a comprehensive review of the block entropy in \cite{Nicolis}. The binary chain is obtained by linearly averaging over 50 realizations, then calculating the spectral densities and applying the dual mapping $0\leftarrow \omega\in(-\infty,0)$ and $1\leftarrow \omega\in(0,\infty)$. 
\begin{figure}[hbt]
	\centering
	\includegraphics[width=0.6\columnwidth]{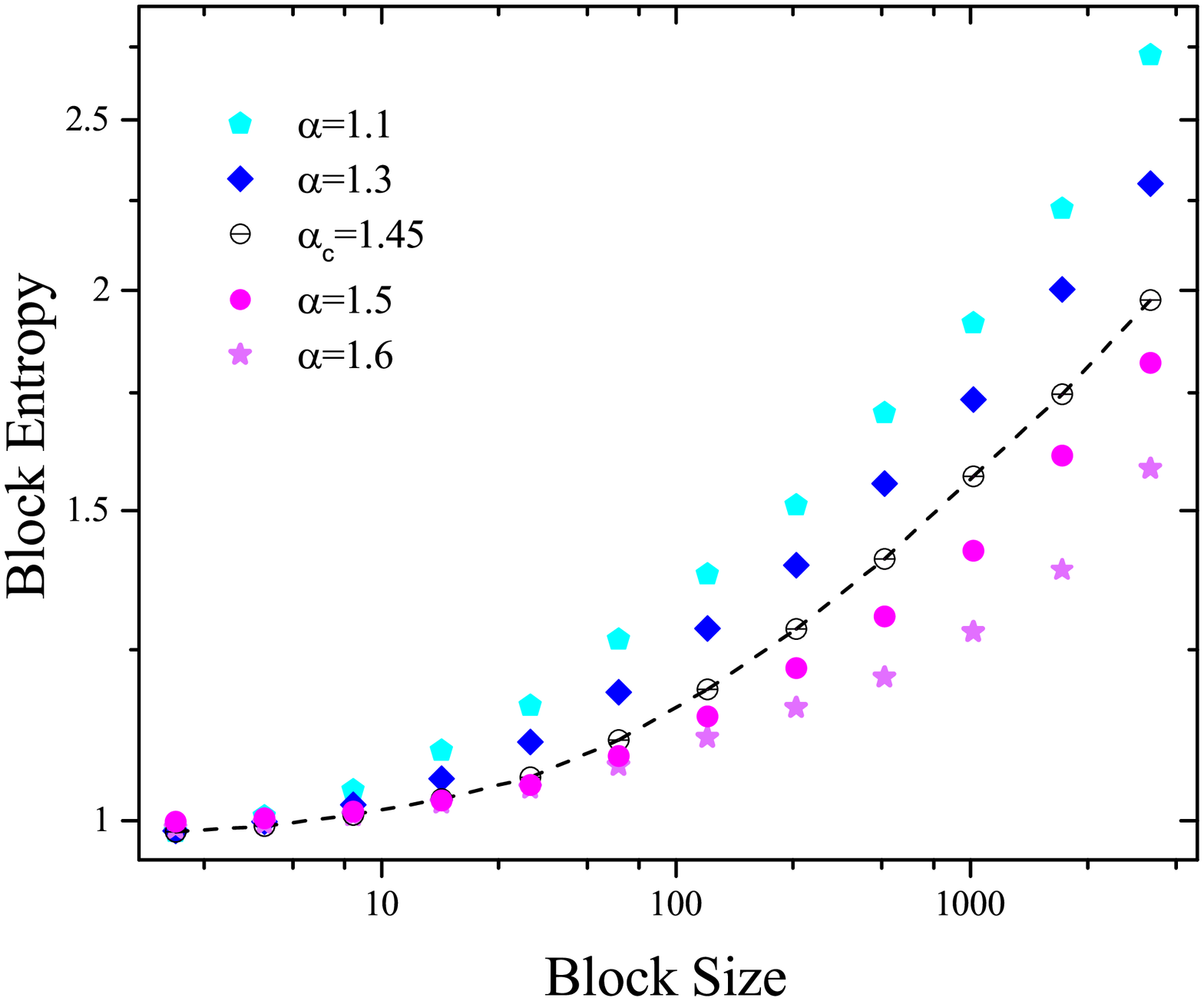}
	\caption{Block entropy vs. the block length $\ell$ for various values of the scaling exponent $\alpha$. With the symbols are plotted the numerical data while the dashed black line denotes the fitting function in Eq.~(\ref{FitFun}). The more ordered the chain structure is ($\alpha>\alpha_c$) the less its entropy value for given block sizes $\ell$.}
	\label{BlockEnt1}
\end{figure}

In Fig.~\ref{BlockEnt1} we recorded LRC time series for five representative values of the parameter $1<\alpha<2$, including the critical value $\alpha_c$. We first want to ensure that the LRC feature of the chains is preserved after the preceding dual mapping of the frequencies.
For this we conduct a quantitative description of the block entropy results by fitting the obtained data with the scaling law proposed in \cite{Nicolis2},
\begin{eqnarray}\label{FitFun}
H_\ell=e+h \ell + g \ell^{\mu}[\ln(\ell)]^{-\nu}\,.
\end{eqnarray}
LRCs are described by the parameter set $\{0<\mu<1,\,\nu>0\}$ \cite{Nicolis2}. Reducing the number of parameters by setting $\mu=\nu$ turned out to be sufficient for our data fitting with $H_\ell$. A representative fit is included in Fig.~\ref{BlockEnt1} for $\alpha=\alpha_c$ (dashed black line). The numerical results are presented  in Table \ref{table1}, from which we verify indeed that $0<\mu<1$. As expected when $1<\alpha<2$, the linear term of $H_\ell$ is negligible ($h\lesssim 10^{-4}$) thereby revealing the dominance of the long-range features in the chains under scrutiny. It is interesting to note that, the value of the exponent $\mu$ of the nonlinear term in $H_\ell$ captures the transition between localized-delocalized bands by exhibiting a maximum for the chain characterized by the critical scaling exponent, $\mu_{\alpha_c}=0.63$.
\begin{table}
	\centering
	\begin{tabular}{ c|c|c|c|c|c }
		$\alpha\;\;\to$& $1.1$ & $1.3$ & $1.45$ & $1.5$ & $1.6$\\	\hline \hline
		\rule{0pt}{15pt}$ e $ & 0.487 & 0.744             &0.917&0.918&0.769\\
		\rule{0pt}{15pt}$ h\times 10^{-5} $ & 8.880 & 3.822 &-21.65  &2.206&4.976 \\
		\rule{0pt}{15pt}$ g $ & 0.393 & 0.172 &0.038&0.050&0.184 \\
		\rule{0pt}{15pt}$\mu=
		\nu$ & 0.246 & 0.344 &0.629 & 0.459 & 0.185\\ \hline
		\rule{0pt}{15pt}$\chi^2\times 10^{-4}$ & 4.90 & 3.10   &0.95 &2.95&0.91
	\end{tabular}
	\caption{Numerical results for the fitting parameters in Eq.~(\ref{FitFun}).}
	\label{table1}
\end{table}
\begin{figure}[hbtp]
	\centering
	\subfigure{\includegraphics[width=0.6\textwidth]{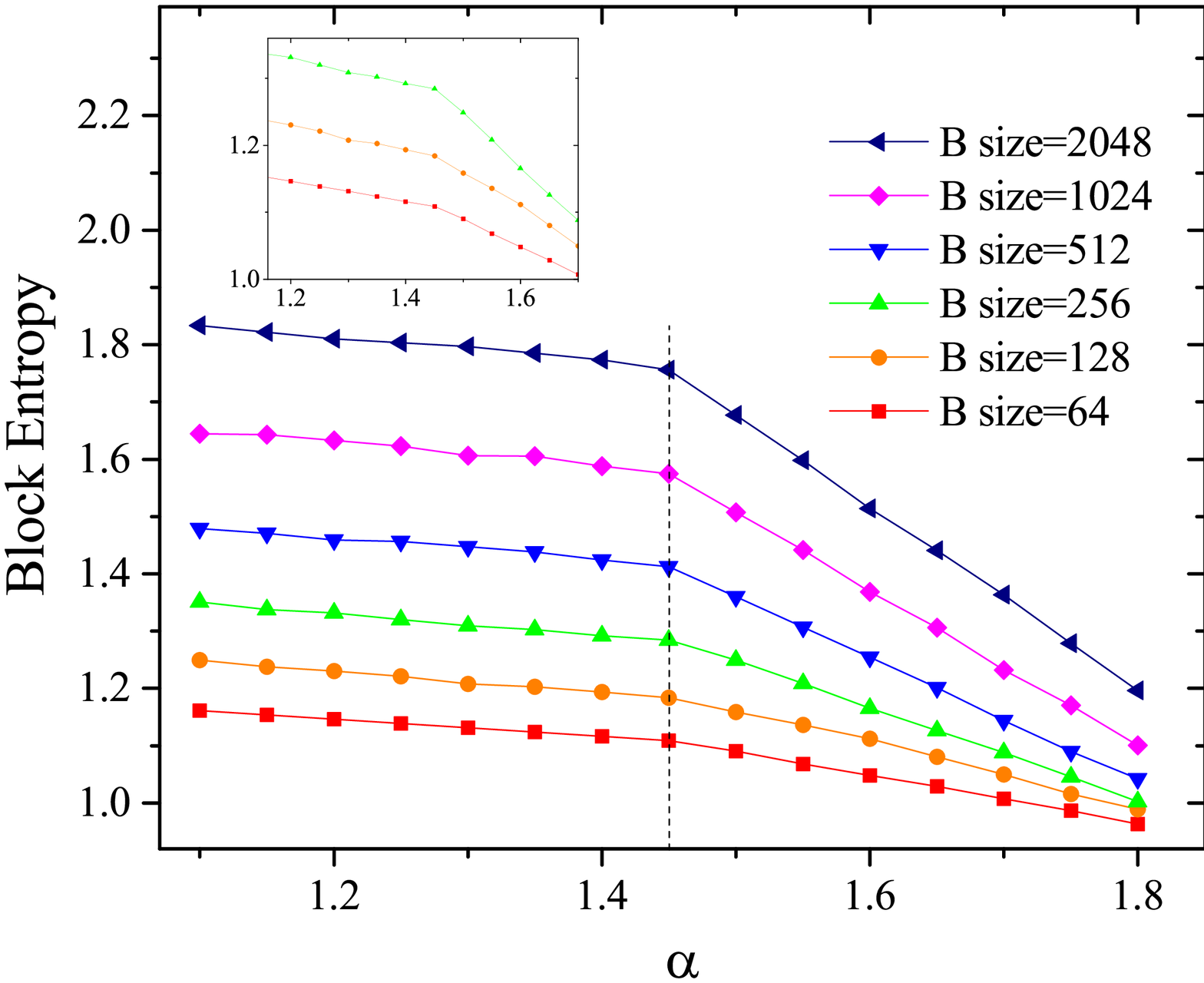}}
	\put(-20,230){$(\mathrm{a})$}
	\quad
	\subfigure{\includegraphics[width=0.6\textwidth]{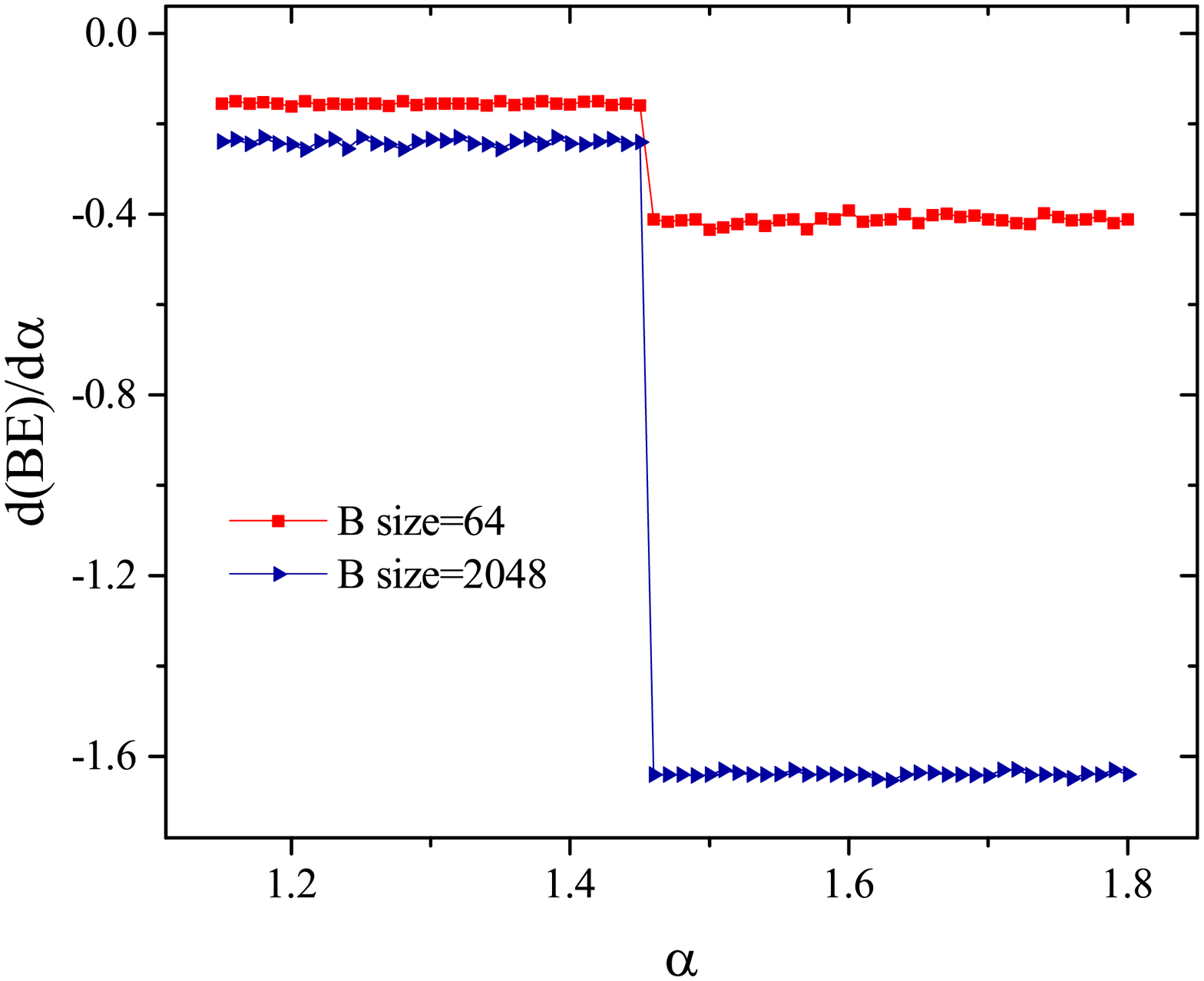}}
	\put(-20,230){$(\mathrm{b})$}
	\caption{\label{fig:alpha2} (a) Plot of the block entropy as a function of the scaling exponent $\alpha$ for different block sizes $\ell=2^n$ and a fixed LRC chain length $N=2^{20}$. The vertical dashed line corresponds to $\alpha_c$. (b) Plot of the rate of change of the block entropy with respect to $\alpha$.}
\end{figure}

Having ensured that our frequency mapping preserves the LRC feature of the chains for $\alpha\in(1,2)$, we now proceed with exploring the behaviour of the block entropy with respect to the scaling exponent $\alpha$ by fixing the block size for each chain. In Fig.~\ref{fig:alpha2}(a) we plot the results of our analysis. Surprisingly, we observe that the block entropy not only correctly detects the critical transition exponent $\alpha_c$ (vertical dashed line), but the qualitative behaviour is analogous to the spectral entropy case. 
Similar to the spectral entropy,  we present in Fig.~\ref{fig:alpha2}(b) the rate of change of the block entropy with $\Delta \alpha=0.01$.

\section{\label{sec:Dis}Summary and discussion}

In this paper we have considered the propagation of elastic waves in a one-dimensional binary chain with different types of correlations in the elasticity distribution. Specifically, we calculated the localization length $\xi$ with respect to the wave frequency $\omega$ for a fixed chain length ($N=2^{18}$). For this we applied the  Transfer Matrix method on the wave equation of interest. The long-range correlated binary chains are parameterized by the scaling exponent $\alpha$. We observed, in agreement with the related literature, that above the critical value of this exponent i.e. $\alpha_c=1.45$, there is an extended delocalization band, while below it, all frequency modes are localized. For the sake of comparison we also studied $n$-mer chains ($n=2,4,6,\ldots$) as representative examples of short-range correlated binary chains. In this case, as shown in \cite{Farzadian2016}, there are $(n-1)$ resonance frequency peaks of delocalized modes. Accordingly, the dimer ($n=2$), is the first $n$-mer exhibiting delocalized modes after a delta-correlated white noise disordered chain where all states are localized.

The concept of statistical entropy as a measure of complexity was invoked to investigate the transition between the localization and delocalization band regimes for long-range correlated sequences. Precisely, we studied two different types of entropy analysis based on the Shannon entropy expression, namely the spectral and the block entropies, as functions of the scaling exponent $\alpha$. We demonstrated that both of them are accurate measures to capture the aforementioned transition. The latter is unveiled by a non-existing entropy derivative at a critical value which coincides with the observed value in  \cite{Carpena2002}, $\alpha_c=1.45$. Even more impressive was the fact that both entropies exhibit the same qualitative behavior, i.e., linear decay with respect to $\alpha$ with different negative slopes around $\alpha_c$. The absolute difference in the slopes was increasing for increasing chain length $N$ or block size $\ell$, respectively.
The relation between all the different values of $\alpha$ and $SE_\alpha$ are consistent with the statistical notion of (dis)order in a chain, i.e., the higher value of $\alpha$ the more ordered the chain structure and the less the entropy.

\section*{Acknowledgements}
\noindent The authors O.F., T.O. and M.G. acknowledge the ORAU grant entitled ``Casimir light as a probe of vacuum fluctuation simplification" with PN 17098 and the state-targeted program ``Center of Excellence for Fundamental and Applied Physics" (BR05236454) by the Ministry of Education and Science of the Republic of Kazakhstan.


\bibliographystyle{model1-num-names}
\bibliography{mybibfile}

\end{document}